\documentclass[12pt]{article}
\usepackage{amssymb,amsmath}
\begin{document}
\title{\textbf{Does massless QCD have vacuum energy?}} 
\author{B. Holdom%
\thanks{bob.holdom@utoronto.ca}\\
\emph{\small Department of Physics, University of Toronto}\\[-1ex]
\emph{\small Toronto ON Canada M5S1A7}}
\date{}
\maketitle
\begin{abstract}
It is widely thought that this question has a positive answer, but we argue that the support for this belief from both experiment and theory is weak or nonexistent. We then list some of the ramifications of a negative answer.
\end{abstract}

There is probably a class of asymptotically free theories of gauged massless fermions that behave similarly to the chiral limit of QCD, and we refer to this class simply as massless QCD. We can consider massless QCD in isolation, defined on a $3+1$ dimensional Minkowski background with no gravity (the path integral representation can probably be defined in a rigorous way \cite{a0}). Massless QCD displays an explicit quantum mechanical breaking of a classical scale invariance. From this trace anomaly and the assumed Lorentz invariance of the vacuum we obtain the vacuum energy density,\footnote{$T$ is the energy momentum tensor, $G$ is the gluon field strength tensor, $\beta$ is the beta function and $\alpha_s=g^2/4\pi$.}
\begin{equation}
\varepsilon=\frac{1}{4} \langle T_\mu^\mu\rangle = \langle\frac{\beta(\alpha_s)}{16\alpha_s} G^a_{\mu\nu}G_a^{\mu\nu}\rangle\label{e1}
.\end{equation}
It is customary to split this last quantity into perturbative and nonperturbative contributions, so that $\varepsilon=\varepsilon_{\rm pert}+\varepsilon_{\rm nonpert}$. Then $\varepsilon_{\rm nonpert}$ is closely related to the so-called gluon condensate, often written simply as $\langle \frac{\alpha_s}{\pi}G^2\rangle$, a quantity that plays a fundamental role in the SVZ sum rule approach to QCD that was developed nearly thirty years ago \cite{a8}. In this context it is conventional \cite{a19} to introduce a scale $\mu$ as a boundary between hard and soft modes, so that only soft modes contribute to the condensate. Then the condensate is $\mu$ dependent, although this dependence is thought to be weak.

\section{Experimental hints}

There have been extensive efforts to determine the gluon condensate experimentally, but to extrapolate these values to massless QCD one must account for the effect of nonvanishing current quark masses. When combined with the chiral symmetry breaking condensate these masses imply a contribution to the gluon condensate determined by a low energy theorem \cite{a9},
\begin{equation}
\frac{d}{dm_f}\langle\frac{\alpha_s}{\pi}G^2\rangle=-\frac{24}{b}\langle\overline{q}_fq_f\rangle,
\end{equation}
where $b=11-\frac{2}{3}n_f$. This holds for small quark masses $m_f$ and if it is assumed to apply to the strange quark then we have the dominant contribution as \cite{a9}
\begin{equation}
\langle\frac{\alpha_s}{\pi}G^2\rangle_{\rm{from\;quark\;masses}}\approx-\frac{24}{b}m_s\langle\overline{s}s\rangle\approx0.006\mbox{ GeV}^4.\label{e2}
\end{equation}
This is of the same sign as the contribution that is expected \cite{a8} to survive in the massless theory. Thus for there to be an experimental indication for the gluon condensate of the massless theory, the experimental value would have to be larger and inconsistent with (\ref{e2}).

Although this may have been true for early determinations of $\langle \frac{\alpha_s}{\pi}G^2\rangle$ this does not appear to be the case at present. We list the following most recent determinations, to be compared with  the original and still popular `standard' value of $\langle \frac{\alpha_s}{\pi}G^2\rangle=0.012$ GeV$^4$ \cite{a8}.
\begin{center}\begin{tabular}{ll}$\tau$ decay (2001) \cite{a1} & $0.006\pm0.012$ GeV$^4$ \\charmonion (vector) \cite{a2} & $0.009\pm0.007$ GeV$^4$ \\charmonion (pseudoscalar) \cite{a3} & $\lesssim 0.008$ GeV$^4$ \\charmonion (axial-vector) \cite{a4} & $0.005^{+0.001}_{-0.004}$ GeV$^4$ \\ valence quark distributions \cite{a5} & $0.006\pm0.006$ GeV$^4$\\$\tau$ decay (2006) \cite{a6} & $0.001\pm0.012$ GeV$^4$\\$\tau$ decay (2008) \cite{a6a} & $-0.015\pm0.008$ GeV$^4$ \end{tabular}
\end{center}
 The first five of these values are also discussed in a 2006 review \cite{a7}. The last result emerges from an analysis that is purported to give the most precise determination of $\alpha_s$ to date. These results clearly offer no support for a nonvanishing gluon condensate in the chiral limit. Our emphasis here is on this lack of support and not that these results are conclusive one way or another. The extraction of any of these results from experiments remains rather indirect and model dependent.

Direct lattice studies of the gluon condensate are also difficult due to the breaking of the spacetime and classical scale symmetries by the lattice regularization. This typically leads to large perturbative contributions that need to be carefully separated from the nonperturbative effects of interest. Also most lattice studies of the gluon condensate are performed in the context of quenched or pure glue theories, but these are not very relevant to the question of vacuum energy in massless QCD. In particular instantons are known to contribute to vacuum energy in the pure glue theory, or in the presence of quarks with Lagrangian level masses \cite{a14}.

Indicative of this lack of relevant results is the fact that we are aware of only one (rather dated) attempted determination of the gluon condensate from a lattice with light dynamical fermions \cite{a10}. The following results for the condensate are reported, where $a$ is the lattice spacing and $m_q$ the quark mass.
\begin{center}\begin{tabular}{cc} $0.031\pm0.005^{+0.012}_{-0.007}$ GeV$^4$ &($am_q=0.02$) \\$0.015\pm0.003^{+0.006}_{-0.003}$ GeV$^4$ &($am_q=0.01$) \end{tabular}
\end{center}
These results appear to be consistent with the condensate vanishing in the $m_q\rightarrow 0$ limit.\footnote{We use the word `appear' since the authors themselves don't draw attention to this.} The study in \cite{a11} also displays a gluon condensate that extrapolates to zero with the quark mass, but this starts from the opposite limit of very heavy quarkonia in NRQCD. And finally, a lattice study \cite{a11a} of the quark propagator found that the contribution of the gluon condensate was negligible in comparison to that of the `mixed' condensate, $\alpha_s\langle\overline{q}(g A\!\!\!/)q\rangle=(-0.11\pm0.03)$ GeV$^4$. Again, we are not claiming that any of these studies are very compelling one way or the other.

We do conclude though that there is a lack of experimental or lattice-based evidence for a nonvanishing gluon condensate in massless QCD. The authors of \cite{a6a}, given that their recent result listed above for the condensate is at variance with the standard value,  note `that not much is known from theoretical grounds about the value of the gluon condensate'. With this perspective we turn to the next section.

\section{Theoretical hints}

We may want to review the theoretical arguments in favor of a nonvanishing condensate. The basic argument is `why not?'. What is there to prevent it? After all, the scale anomaly implies dimensional transmutation and thus a mass scale $\Lambda_{\rm QCD}$, which in turn sets the scale of the chiral symmetry breaking quark condensate. Shouldn't it also enter the gluon condensate? And if the gluon condensate is defined in the conventional way \cite{a19} then it is $\mu$ dependent in which case it cannot identically vanish.

Our main interest is in the complete vacuum energy density $\varepsilon=\varepsilon_{\rm pert}+\varepsilon_{\rm nonpert}$. The perturbative contribution faces its own problems of definition; it is sensitive to the regulator used. We thus temporarily turn to a \textit{free} massless theory, for which there is also an apparent infinite zero point energy. But we notice that if we consider the analog of (\ref{e1}) for a free theory we would conclude that the vacuum energy vanishes. In other words the combination of Lorentz invariance and scale invariance implies vanishing vacuum energy. Is there a regulator that can make sense of this? Yes; dimensional regularization gives $\langle T_{\mu\nu}\rangle=0$ directly, due to the `identity'
\begin{equation}
\int dk\,k^\alpha =0\mbox{ for }\alpha>-1.
\label{e3}\end{equation}
This is a statement of dimensional analysis and the lack of any mass scale. As discussed in \cite{a12}, other regulators, including cutoffs of the 3 or 4 dimensional variety, cannot be trusted for the calculation of $\langle T_{\mu\nu}\rangle$. Such regulators do not respect the scale invariance, and they end up violating Lorentz invariance by producing $ 0\neq\langle T_{\mu\nu}\rangle\propto\!\!\!\!\!/\;\eta_{\mu\nu}$.\footnote{Normal ordering in the free theory also removes the vacuum energy, doing so even when masses are present. Dimensional regularization more sensibly gives nonvanishing vacuum energy in a free massive theory, consistent with broken scale invariance.}

Given these observations for a free theory, we are thus led to apply dimensional regularization to the calculation of $\varepsilon_{\rm pert}$ in massless QCD. A direct calculation involves the insertion of $T_\mu^\mu=(d-4)G^a_{\mu\nu}G_a^{\mu\nu}/4$ (the classical relation where $d$ is the spacetime dimension) into vacuum diagrams, where in principle loop effects can cancel the $d-4$ dependence. This is certainly true when there are current quark masses, but in our massless case the result $\varepsilon_{\rm pert}=0$ persists. A loop integration in each vacuum diagram can always be reduced \cite{a15} to the form (\ref{e3}) or
\begin{equation}
\int dk\,k^\alpha(\log k)^\beta(\log\log k)^\gamma...=0\mbox{ for }\alpha>-1.
\label{e4}\end{equation}
Dimensional regularization extends the dimensional analysis argument for the vanishing of vacuum diagrams in massless theories to all orders in perturbation theory.

This result is consistent with the scale anomaly, whose expectation value appears in (\ref{e1}) and whose perturbative existence is due to the ultraviolet behavior of loops. Basically the scale anomaly is an operator statement, and the renormalization of the $G^a_{\mu\nu}G_a^{\mu\nu}$ operator cancels the $d-4$ dependence in the classical relation. This is related to the introduction of a renormalization mass scale. It is similar for the background field calculation of the scale anomaly in dimensional regularization \cite{a13}; here the background field provides an external momentum which ensures a nonvanishing result. Thus dimensional regularization unites the explicit breaking of scale invariance with the vanishing of the perturbative vacuum energy in a massless theory.  Dimensional regularization, although not necessarily unique in this regard, has been emphasized before because of this property \cite{a15}.

If we quell any further qualms about the physical significance, or lack thereof, of regularization schemes we can proceed. The nonperturbative vacuum energy density $\varepsilon_{\rm nonpert}$ is now well defined and unambiguous, since it constitutes the complete result. The true vacuum deviates significantly from the perturbative vacuum in the infrared, where the coupling is large and where the propagating degrees of freedom have completely changed. The resulting $\varepsilon=\varepsilon_{\rm nonpert}=\frac{1}{4}\langle T_\mu^\mu\rangle$ should be characterized by these energy scales and thus presumably is of order $\Lambda_{\rm QCD}^4$. If this is the case then nonperturbative effects have translated the explicit breaking of scale invariance into vacuum energy. This is known to occur in gauge theories \textit{without} massless fermions, since then instantons induce a vacuum energy dependent on a vacuum angle $\theta$ \cite{a14}. 

When massless fermions are present, as in massless QCD, then configurations with net winding number do not contribute to vacuum energy. Note that contrary to intuition, the $\theta$ dependence disappears even in the presence of the chiral condensate. Other possible contributions to vacuum energy include instanton-anti-instanton configurations and effects coming purely from the chiral condensate. Effective actions are often used to study such effects, in particular the nonlocal actions \cite{a13a} and other dynamical models used to study chiral symmetry breaking. It is evident that as soon as dimensionful quantities appear in the effective equations of motion of some effective description, then a vacuum energy is apparently generated. But these effective descriptions are ambiguous with respect to an overall additive constant in the action. This constant should be chosen so that the $\langle T_\mu^\mu\rangle$ as calculated in the effective theory matches that of the underlying theory (i.e.~adjusted order by order in the perturbative expansion of the effective action). Thus while massive effective theory descriptions are useful for many purposes, they shed little light on the vacuum energy of the complete theory.

In massless QCD we may consider a sufficient number of massless fermions so that the theory has a perturbative Banks-Zaks \cite{a18} fixed point. The theory flows from a free, scale-invariant theory in the ultraviolet to a weakly interacting, scale-invariant theory in the infrared. Since $\langle T_\mu^\mu\rangle=0$ in perturbation theory, it could be expected that this is the complete answer in this case. We may then have an example of a theory that contains explicit breaking of scale invariance and yet has vanishing vacuum energy.

As the number of fermions is reduced in the Banks-Zaks construction, the infrared fixed point moves to stronger coupling. Eventually the coupling is sufficiently large for chiral symmetry breaking. The fermions decouple and the antiscreening nature of the gluons continues to push the gauge coupling to arbitrarily large values. This is the old picture of `infrared slavery'.  The implied singular behavior of the coupling at some scale $\Lambda_{\rm QCD}$ reinforces the belief that $\langle T_\mu^\mu\rangle\neq0$. Indeed, the infrared Landau pole and the associated `infrared renormalons' have served as a primary argument for the existence of a gluon condensate  \cite{a20, a19}. This approach attempts to deduce nonperturbative aspects of the theory from its perturbative properties. 

But more direct attempts to extract nonperturbative information portray a different picture. An analysis of the Schwinger-Dyson equations of the QCD Green's functions in Landau gauge indicates the existence of a nontrivial infrared fixed point \cite{a16}. Here ghost-gluon, triple-gluon and quartic-gluon couplings are found to approach constant values, as also suggested by earlier work \cite{a21} and renormalization group methods \cite{a22}. In this picture massless QCD flows from a free, scale-invariant theory in the ultraviolet to a strongly interacting scale-invariant theory in the infrared. $\Lambda_{\rm QCD}$ characterizes the cross-over region. Confinement remains to be fully understood and the latest lattice studies in Landau gauge have not confirmed all aspects of this picture \cite{a25} (see also \cite{a25a}). But all these studies agree that the Landau pole of naive perturbation theory is replaced by more smooth behavior.

The discussion in this section does not prove the absence of vacuum energy, since in the end this is a question of nonperturbative physics. But our discussion does point to a lack of any rigorous theoretical argument for nonvanishing vacuum energy in massless QCD. The possible vanishing of $\langle T_\mu^\mu\rangle$ in the presence of the scale anomaly $T_\mu^\mu\neq0$ can be viewed as a surviving remnant of the original classical scale invariance. Expressed as a property of the full generating functional $W$ it is $\delta W/\delta\lambda|_{\lambda=1}=0$, where the variation is with respect to a global rescaling of the background spacetime metric, $\eta_{\mu\nu}\rightarrow\lambda\eta_{\mu\nu}$.

\section{Implications}

The question of vacuum energy in massless QCD is of more than academic interest. A vanishing vacuum energy, if true, would indeed have some interesting consequences.

1) It would breath new life into the idea of induced gravity \cite{a17,a15}. In this picture the metric of a Lorentzian manifold is treated as a background field for which an effective gravitational action is generated through the quantum effects of matter. We choose matter to be described by a theory of the massless QCD type with an adjustable $\Lambda_{\rm QCD}$. In particular the induced Newton's constant $G$ is given by \cite{a15a}
\begin{equation}
\frac{1}{G}=i\frac{\pi}{6}\int d^4x\,x^2\langle{\cal T}(\hat{T}(x)\hat{T}(0))\rangle,\quad\hat{T}=T_\mu^\mu-\langle T_\mu^\mu\rangle.
\end{equation}
The perturbative contributions are quadratically divergent and thus vanish in massless QCD via dimensional regularization as before \cite{a15}. But for this quantity nonperturbative effects can be expected to make a contribution of order $\Lambda_{\rm QCD}^2$, which can be adjusted to yield the correct $1/G$. Concurrently a vanishing vacuum energy, $\langle T_\mu^\mu\rangle=0$, translates into the absence of a cosmological constant in the otherwise general gravitational action that results. And although this effective action explicitly breaks scale invariance, it vanishes for the $g_{\mu\nu}=\eta_{\mu\nu}$ vacuum solution, and in this way it captures the invariance $\eta_{\mu\nu}\rightarrow\lambda\eta_{\mu\nu}$.

2) It would motivate the search for a complete theory of particle physics based on an asymptotically free massless theory of gauged fermions. This would most likely be a chiral gauge theory, and in a sequential breakdown it would have to generate the electroweak scale and above this a flavor scale, in addition to the Planck scale. All masses have a dynamical origin, with the small values of quark and lepton masses arising due to a ratio of dynamical mass scales. Along with the natural generation of large hierarchies and the inducement of gravitation, the additional intriguing possibility is a vanishing vacuum energy, related as we have described to the classical scale symmetry of the complete theory.

3) Effective field theory would still provide useful descriptions of particular physics on whatever energy scales of interest. These theories may well incorporate cutoffs and massive particles, thus hiding in particular the ultimate dynamical origin of quark and lepton masses. An appropriate cosmological constant may have to appear in any such effective theory to ensure that the total vacuum energy vanishes. But this is not an unnatural tuning, it is the result of a matching to the underlying theory.

4) The LHC may uncover a dynamical mechanism for electroweak symmetry breaking. Then the measurements of the gluon condensate of QCD will turn out to have been a foreshadowing of this discovery.

\section*{Acknowledgments}
I thank Erich Poppitz and Michael Luke for interesting discussions. This work was supported in part by the National Science and Engineering Research Council of Canada.


\begin{thebibliography}{11}
\bibitem{a0} G.~`t Hooft, ``On peculiarities and pit falls in path integrals``, hep-th/0208054.
\bibitem{a8} M.~Shifman, A.~Vainshtein and V.~Zakharov, Nucl.~Phys.~B147 (1979) 385; 448.
\bibitem{a19} S.~Shifman, Lecture given at the 1997 Yukawa International Seminar Non-Perturbative QCD-Structure of the QCD Vacuum, Kyoto, hep-ph/9802214, and references therein.
\bibitem{a9} V.A.~Novikov, M.~Shifman, A.~Vainshtein and V.~Zakharov, Nucl.~Phys.~B191 (1981) 301 (section 14).
\bibitem{a1} B.V.~Geshkenbein, B.L.~Ioffe and K.N.~Zyablyuk, Phys.~Rev.~D 64 (2001) 093009.
\bibitem{a2} B.L.~Ioffe and K.N.~Zyablyuk, Eur.~Phys.~J.~C 27 (2003) 229.
\bibitem{a3} K.N.~Zyablyuk, JHEP 0301:081 (2003).
\bibitem{a4} A.V Samsonov, hep-ph/0407199.
\bibitem{a5} B.L.~Ioffe, A.G.~Oganesian, Nucl.~Phys.~A 714 (2003) 145
\bibitem{a6} M.~Davier, A.~Hocker, and Z.~Zhang, Rev.~Mod.~Phys., 78 (2006). 
\bibitem{a6a} M.~Davier, S.~Descotes-Genon, A.~Hocker, B.~Malaescu, and Z.~Zhang, arXiv:0803.0979.
\bibitem{a7} B.L.~Ioffe, Prog.~Part.~Nucl.~Phys.~56 (2006) 232.
\bibitem{a14} G.~`t Hooft, Phys.~Rev.~D14 (1976) 3432.
\bibitem{a10} M.~D'Elia, A.~Di Giacomo, and E.~Meggiolaro, Phys.~Lett.~B 408 (1997) 315, hep-lat/9705032.
\bibitem{a11} J.~Fingberg, Nucl.Phys.Proc.Suppl.73:348-350,1999, hep-lat/9810050.
\bibitem{a11a} E.R.~Arriola, P.O.~Bowman, W.~Broniowski, Phys.~Rev.~D70 (2004) 097505, hep-ph/0408309.
\bibitem{a12} E.K.~Akhmedov, ``Vacuum energy and relativistic invariance'', hep-th/0204048.
\bibitem{a15} S.~Adler, Rev.~Mod.~Phys.~54 (1982) 729.
\bibitem{a13} M.~Peskin and D.~Schroeder, ``An Introduction to Quantum Field Theory'', Addison Wesley.
\bibitem{a13a} J.M.~Cornwall, R.~Jackiw and E.~Tomboulis, Phys.~Rev.~D10 (1974) 2428.
\bibitem{a18} T.~Banks and A.~Zaks, Nucl.~Phys.~B196 (1982) 189.
\bibitem{a20} A.~H.~Mueller, in Proc.~Int.~Conf.~QCD Ð 20 Years Later, Aachen 1992, eds.~P.~Zerwas and H.~Kastrup, (World Scientific, Singapore, 1993), vol.~1, page 162.
\bibitem{a16} R.~Alkofer, C.S.~Fischer, and F.J.~Llanes-Estrada, Phys.~Lett.~B611 (2005) 279, hep-th/0412330; C.S.~Fischer, J.~Phys.~G32 (2006) R253, hep-ph/0605173; C.S.~Fischer, J.M.~Pawlowski, Phys.~Rev.~D75 (2007) 025012, hep-th/0609009.
\bibitem{a21} L.~von Smekal, A. Hauck, and R. Alkofer, Ann. Phys. 267 (1998) 1, hep-ph/9707327; C.~Lerche and L.~von Smekal Phys.~Rev.~D65 (2002) 125006, hep-ph/0202194.
\bibitem{a22} J.M.~Pawlowski, D.F.~Litim, S.~Nedelko, and L.~von Smekal, Phys.~Rev.~Lett.~93 (2004) 152002, hep-th/0312324; C.S.~Fischer and H.~Gies, JHEP 10 (2004) 048, hep-ph/0408089; H.~Gies, Phys.~Rev.~D66 (2002) 025006, hep-th/0202207.
\bibitem{a25} A.~Cucchieri and T.~Mendes, arXiv:0710.0412;  I.L.~Bogolubsky, E.M.~Ilgenfritz, M.~Muller-Preussker and A.~Sternbeck, arXiv:0710.1968; A.~Sternbeck, D.B.~Leinweber, L.~von Smekal and A.G.~Williams, arXiv:0710.1982.
\bibitem{a25a} Ph.~Boucaud, J.P.~Leroy, A.~LeYaouanc, A.Y.~Lokhov, 
J.~Micheli, O.~Pene, J.~Rodrõguez-Quintero and C.~Roiesnel, hep-ph/0702092.
\bibitem{a24} V.N.~Gribov, Nucl.~Phys.~B139 (1978) 1.
\bibitem{a17} A.D.~Sakharov, Sov.~Phys.~Dokl.~12 (1968) 1040; Y.B.~Zel'dovich, JETP Lett 6 (1967) 316. 
\bibitem{a15a} S.~Adler, Phys.~Lett.~B95 (1980) 241; A.~Zee, Phys.~Rev.~D23 (1981) 858.

\end{thebibliography}
\end{document}